\documentclass[aps,prl,twocolumn,superscriptaddress,showpacs,floatfix,amsmath,amssymb,nofootinbib,preprintnumbers]{revtex4-1}

\usepackage{graphicx}

\usepackage{epstopdf}
\usepackage{color}

\newcommand{\be}{\begin{equation}}
\newcommand{\ee}{\end{equation}}
\newcommand{\ba}{\begin{eqnarray}}
\newcommand{\ea}{\end{eqnarray}}

\newcommand{\beq}{\begin{equation}}
\newcommand{\eeq}{\end{equation}}
\newcommand{\beqa}{\begin{eqnarray}}
\newcommand{\eeqa}{\end{eqnarray}}

\begin{document}

\title{Reentrant Phase Transitions in Rotating AdS Black Holes}
\author{Natacha Altamirano}
\email{naltamirano@perimeterinstitute.ca}
\affiliation{Perimeter Institute, 31 Caroline St. N. Waterloo
Ontario, N2L 2Y5, Canada}
\affiliation{Department of Physics and Astronomy, University of Waterloo,
Waterloo, Ontario, Canada, N2L 3G1}
\author{David Kubiz\v n\'ak}
\email{dkubiznak@perimeterinstitute.ca}
\affiliation{Perimeter Institute, 31 Caroline St. N. Waterloo
Ontario, N2L 2Y5, Canada}
\author{Robert B. Mann}
\email{rbmann@uwaterloo.ca}
\affiliation{Perimeter Institute, 31 Caroline St. N. Waterloo
Ontario, N2L 2Y5, Canada}
\affiliation{Department of Physics and Astronomy, University of Waterloo,
Waterloo, Ontario, Canada, N2L 3G1}

\date{June 24, 2013}  % revised version
%\date{\today}

\begin{abstract}
We  study the thermodynamics of higher-dimensional singly spinning asymptotically AdS black holes in 
the canonical (fixed $J$) ensemble of extended phase space, where the cosmological constant 
is treated as pressure and 
the corresponding conjugate quantity is interpreted as thermodynamic volume.  Along with the usual 
small/large black hole phase transition,  we find a new phenomenon of
reentrant phase transitions for   all $d\geq 6$ dimensions, in which a monotonic variation of the temperature yields two phase transitions from large to small and back to large black holes.  This situation is similar to that seen in multicomponent liquids.
\end{abstract}

%\preprint{DAMTP-2011-80}
% 04.50.-h  Higher-dimensional gravity and other theories of gravity
% 04.50.Gh  Higher-dimensional black holes, black strings, and related objects
% 04.70.Bw  Classical black holes
% 04.20.Jb  Exact solutions
% 05.70.Ce Thermodynamic functions and equations of state
% 04.70.-s Physics of black holes
\preprint{pi-stronggrv-332}

\maketitle

%%%%%%%%%%%%%%%%%%%%%%%%%%%%%%%%%%%%%%%%%%%%%%%%%%%%%%%%%
%%%%%%%%%%%%%%%%%%%%%%%%%%%%%%%%%%%%%%%%%%%%%%%%%%%%%%%%%
%\section{Introduction}
{\em Introduction.} 
In view of the AdS/CFT correspondence, phase transitions in asymptotically AdS black holes 
allow for a dual interpretation in the thermal conformal field theory (CFT) living on the AdS boundary---the principal example being the well known  radiation/Schwarzschild-AdS black hole Hawking--Page transition \cite{HawkingPage:1983}   
which can be interpreted as a confinement/deconfinement phase transition in the dual quark gluon plasma \cite{Witten:1998b}.
Charged \cite{ChamblinEtal:1999a,ChamblinEtal:1999b,CveticGubser:1999a, Johnson:2013} and rotating 
\cite{CaldarelliEtal:2000, TsaiEtal:2012} asymptotically AdS back holes possess an interesting 
feature---they allow for a first order small-black-hole/large-black-hole phase (SBH/LBH) transition which is in many ways reminiscent of the liquid/gas transition of the Van der Waals fluid.
This superficial analogy was recently found more intriguing \cite{KubiznakMann:2012} by considering a thermodynamic analysis in an extended phase space where the cosmological constant is identified with thermodynamic pressure and its variations are included in the first law of black hole thermodynamics.  This notion emerges from geometric derivations of the Smarr formula \cite{KastorEtal:2009} that  i) imply the  mass of an AdS black hole should be interpreted as the enthalpy of the spacetime and ii) allow for a computation of the conjugate thermodynamic volume.  Intensive and extensive quantities are now properly identified \cite{KubiznakMann:2012}  and the SBH/LBH transition can be understood as a liquid/gas phase transition by employing Maxwell's equal area law to the $P-V$ diagram. Coexistence lines and critical exponents are then seen to match those of a Van der Waals fluid.

In this paper we report the finding of an interesting phenomena, observed previously in multicomponent fluids, e.g., \cite{NarayananKumar:1994}, of {\it black hole reentrant phase transitions} (RPTs).  A system undergoes an RPT if a monotonic variation of any thermodynamic quantity results in two (or more) phase transitions such that the final state is macroscopically similar to the initial state.  We find for a certain range of pressures (and a given angular momentum) that a monotonic lowering of the temperature yields a large-small-large black hole transition, where we refer to the latter `large' state as an intermediate black hole (IBH).  This situation is accompanied by a discontinuity in  the global minimum of the Gibbs free energy, referred to
as a {\it zeroth-order phase transition}, a phenomenon seen in superfluidity and superconductivity  \cite{Maslov:2004}, and recently for Born--Infeld black holes \cite{GunasekaranEtal:2012}.  We find the RPT to be generic  for all rotating AdS black holes in $d\geq 6$ dimensions.

%\section{Rotating AdS black holes} 
 
%\subsection{Thermodynamics in extended phase space}  
{\em Extended phase space thermodynamics.}
Rotating AdS black holes were constructed in $d=4$ by Carter \cite{Carter:1968cmp} and later generalized to all higher dimensions  \cite{HawkingEtal:1999, GibbonsEtal:2004, GibbonsEtal:2005}. In what follows we limit ourselves to the case of singly spinning black holes for which only one of the rotation parameters is non-trivial. In $d$ spacetime dimensions the metric reads 
\ba
ds^2&=&-\frac{\Delta}{\rho^2}(dt-\frac{a}{\Xi}\sin^2\!\theta d\varphi)^2+
\frac{\rho^2}{\Delta}dr^2+\frac{\rho^2}{\Sigma}d\theta^2\nonumber\\
&+&
\frac{\Sigma \sin^2\!\theta}{\rho^2}[adt-\frac{(r^2+a^2)}{\Xi}d\varphi]^2+r^2\cos^2\!\theta d\Omega_{d-2}^2\,,\qquad
\ea
where $d\Omega_{d-2}^2$ is the metric for the $(d-2)$-sphere and
\ba
\Delta&=&(r^2+a^2)(1+\frac{r^2}{l^2})-2mr^{5-d}\,,\quad \Sigma=1-\frac{a^2}{l^2}\cos^2\!\theta\,,\nonumber\\
\Xi&=&1-\frac{a^2}{l^2}\,,\quad \rho^2=r^2+a^2\cos^2\!\theta\,,
\ea
with $l$ the AdS radius. 
The associated thermodynamic quantities  read (in Planck units) \cite{GibbonsEtal:2005}
\begin{eqnarray}
M&=&\frac{\omega_{d-2}}{4\pi}\frac{m}{\Xi^2}\left(1+\frac{(d-4)\Xi}{2}\right)\,, \label{BHOM} \\
J&=&\frac{\omega_{d-2}}{4 \pi}\frac{ma}{\Xi^2}\,,\quad  {\Omega}_H=\frac{a}{l^2}\frac{r_+^2+l^2}{r_+^2+a^2}\,,\quad
\label{BHJ}\\
T&=&\frac{1}{2\pi}\Bigr[r_+\Bigl(\frac{r_+^2}{l^2}+1\Bigr) \left(\frac{1}{a^2+r_+^2}+
\frac{d-3}{2 r_+^2}\right)-\frac{1}{r_+}\Bigr]\,,\qquad \label{BHT} \\
S&=&\frac{\omega_{d-2}}{4}\frac{(a^2+r_+^2) r_+^{d-4}}{\Xi}=\frac{A}{4}\,, \label{BHS}
\end{eqnarray} 
where $r_+$ is the black hole horizon radius (the largest positive real root of $\Delta=0$)
and $\omega_{d}=2\pi^{\frac{d+1}{2}}/\Gamma[(d+1)/2]$ is the volume of the unit $d$-sphere. 

We interpret the negative cosmological constant $\Lambda$ as a positive thermodynamic 
pressure $P$
 \cite{CaldarelliEtal:2000, KastorEtal:2009, CveticEtal:2010, Dolan:2012}  
 \be\label{PLambda}
P = - \frac{1}{8 \pi} \Lambda=\frac{(d-1)(d-2)}{16 \pi l^2}\, ,
\ee
in which case the  first law of black hole thermodynamics and Smarr formula \cite{KastorEtal:2009} are
\ba\label{1st}
\delta M &=& T\delta S+\Omega_H \delta J+V\delta P\,, \\
\frac{d-3}{d-2}M &=&TS+\Omega_H J-\frac{2}{d-2} VP\,, \label{Smarr}
\ea
where the thermodynamic volume conjugate  to $P$ is \cite{CveticEtal:2010}
\be\label{VBH}
V=\frac{r_+A}{d-1}\Bigl[1+\frac{a^2}{\Xi}\frac{1+r_+^2/l^2}{(d-2)r_+^2}\Bigr]\,.
\ee
It obeys the so called reverse isoperimetric inequality
 \cite{CveticEtal:2010} and in the non-rotating case becomes $V=\frac{\omega_{d-2} r_+^{d-1}}{d-1}\,,$
 which is the spatial volume of a round sphere of radius $r_+$ in the Euclidean space.

%\subsection{Zeroth order phase transition}

{\em Reentrant phase transition.}
The thermodynamic behavior of the system is governed by the Gibbs free energy $G=G(T,P,\dots)$, which reads ($I$ being the Euclidean 
action \cite{GibbonsEtal:2005})
\ba
G\!&=&\!M-TS=\frac{I}{\beta}+\Omega_H J \\
\!&=&\!\frac{\omega_{d-2}r_+^{d-5}}{16\pi \Xi^2}\Bigl(3a^2+r_+^2-\frac{(r_+^2\!-\!a^2)^2}{l^2}+\frac{3a^2r_+^4+a^4r_+^2}{l^4}\Bigr)
\nonumber
\ea
and  depends on an external parameter $J$. We can plot it for fixed $J$ parametrically, first expressing $a=a(J,r_+, P)$ using \eqref{BHJ} and then inserting the (well-behaved) solution into the expressions for $G$ and $T$, consequently expressed  as functions of $P$ and $r_+$.  

A simple criterion used for investigating the thermodynamic stability is the positivity of the specific heat. 
For a canonical ensemble in the extended phase space, it is natural to consider the specific heat at constant pressure
\begin{equation}\label{CP}
 C_P=T\left(\frac{dS}{dT}\right)_P\,
\end{equation}
which is different from the specific heat at constant thermodynamic volume, $C_V$.
We take negativity of $C_P$ as a sign of local thermodynamic instability. 
Note that, as always, we calculate this quantity for fixed $J$. That is, our specific heat at constant $P$
is in fact a specific heat at constant $(P,J)$ and coincides with $C_J$ considered in previous studies, e.g. \cite{MonteiroEtal:2009}. When plotting the Gibbs free energy we plot branches  with $C_P>0$ in red solid lines and
branches with $C_P<0$ in dashed blue lines.

\begin{figure}
\begin{center}
%\rotatebox{-90}{
\includegraphics[width=0.39\textwidth,height=0.25\textheight]{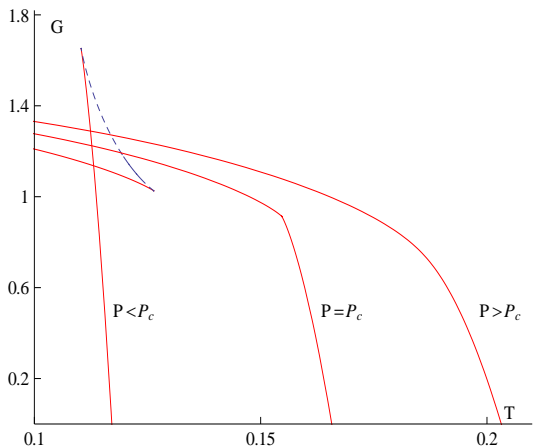}
%}
\caption{{\bf Gibbs free energy in $d=5$ for various values of $P$ and $J=1$.}
As with RN-AdS black holes in any dimension, we see characteristic swallowtail behaviour  indicating an SBH/LBH transition. Solid-red/dashed-blue lines correspond to $C_P$ positive/negative respectively; the  $C_P<0$  line indicates a local thermodynamic instability  where  the Gibbs energy is not a local  minimum; at the joins  $C_P$  diverges. The behaviour of $G$ in $d=4$ is similar. 
}  
\label{Fig:5DKerr}
\end{center}
\end{figure}

The behavior of $G$ depends crucially on the dimension $d$. For $d=4$ and $d=5$ the situation is illustrated in fig.~\ref{Fig:5DKerr}.  For $P>P_c$ and any temperature there is only one branch of locally thermodynamically stable black holes (with positive $C_P$) whereas for  $P<P_c$  the characteristic swallowtail behaviour indicating the SBH/LBH phase transition emerges,
with the global minimum of $G$ having $C_P>0$.  The corresponding $P-T$ diagram (not shown) is reminiscent of what was observed for charged black holes in \cite{KubiznakMann:2012} and is analogous to the Van der Waals $P-T$ diagram. 

\begin{figure}
\begin{center}
%\rotatebox{-90}{
\includegraphics[width=0.39\textwidth,height=0.25\textheight]{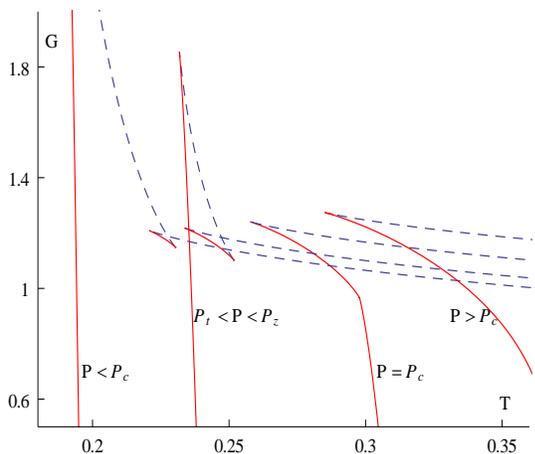}
%}
\caption{{\bf Gibbs free energy in $d=6$ for various values of $P$ and $J=1$.} 
Solid-red/dashed-blue lines correspond to
$C_P$ positive/negative respectively.  As with Schwarzschild-AdS black holes, for $P\geq P_c$, the (lower) LBH branch is thermodynamically stable  whereas the upper branch is unstable.  For $P=P_c$ we observe critical behaviour. At the joins of dashed (blue) and solid (red) lines  $C_P$ diverges. For  $P\in(P_t,P_z)$ we observe a
``zeroth-order phase transition'' signifying the onset of an RPT.}
\label{Fig:6DKerr}
\end{center}
\end{figure}

\begin{figure}
\begin{center}
%\rotatebox{-90}{
\includegraphics[width=0.39\textwidth,height=0.25\textheight]{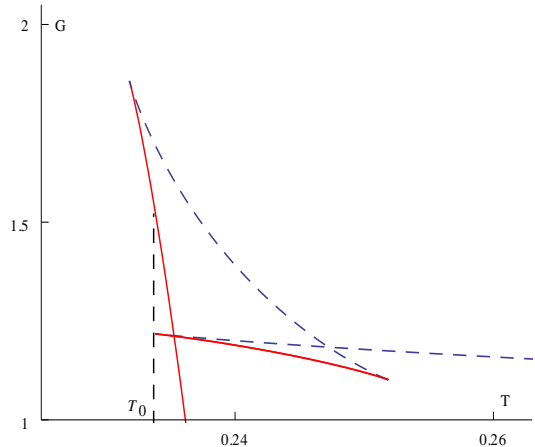}
%}
\caption{{\bf Zeroth order phase transition in $d=6$.}
A close-up of fig. \ref{Fig:6DKerr} illustrating the discontinuity in the global minimum of $G$ at $T=T_0\approx 0.2339\in(T_t,T_z)$
(denoted by the vertical line).
 We have set $P=0.0564\in(P_t,P_z)$ and $J=1$.  Solid-red/dashed-blue lines correspond to $C_P$ positive/negative respectively.}  
\label{Fig:6DKerrZorder}
\end{center}
\end{figure}

\begin{figure}
\vspace{-0.7cm}
\begin{center}
\includegraphics[width=0.47\textwidth,height=0.32\textheight]{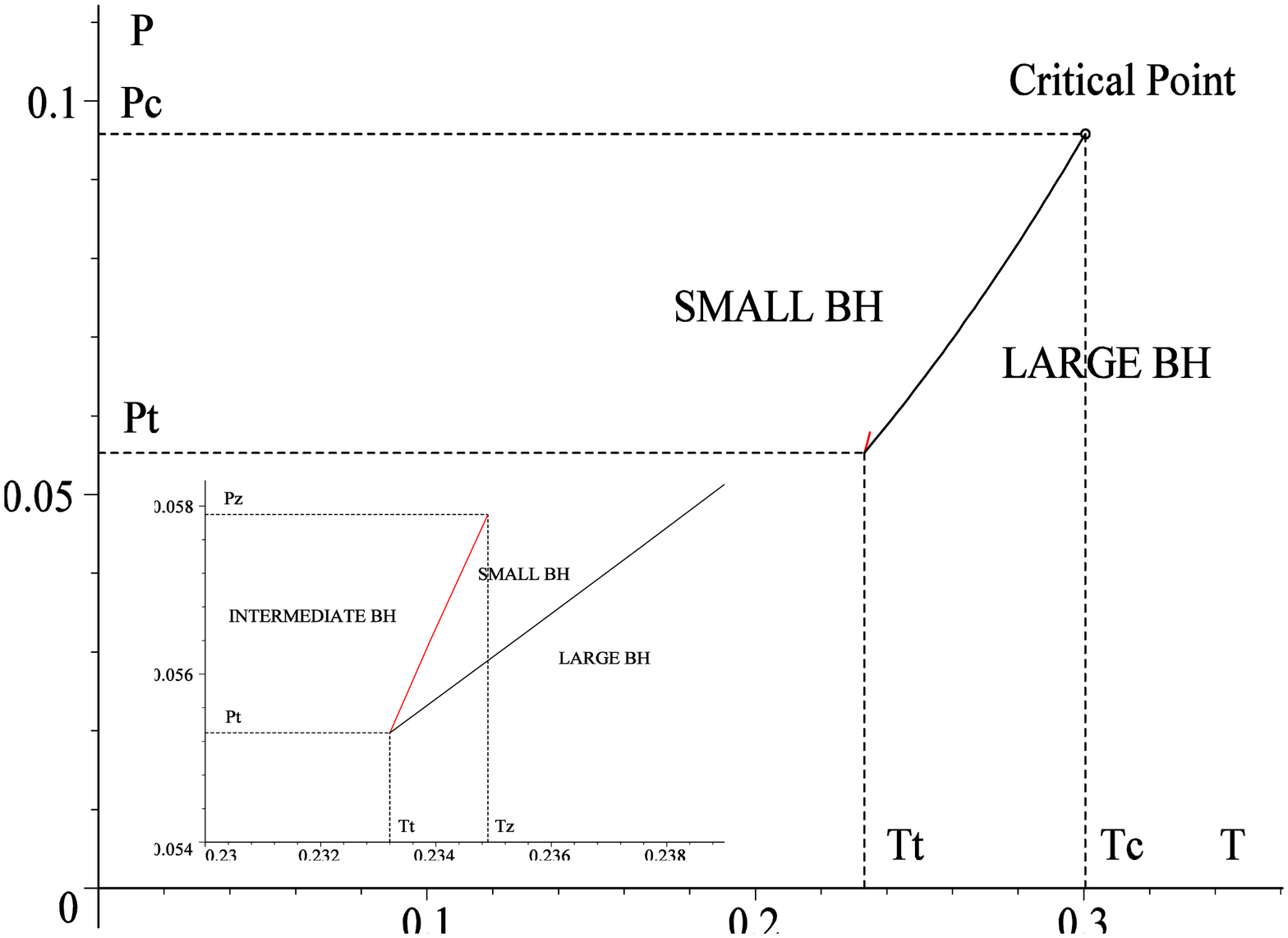}
\caption{{\bf $P-T$ diagram in $d=6$.}
The coexistence line of the first order phase transition between small and large black holes is depicted by 
a thick black solid line for $J=1$. It initiates from the critical point $(P_c, T_c)$ and terminates at $(P_t, T_t)$. The red solid line in the inset indicates the `coexistence line' of small and intermediate black holes, separated by a finite gap in $G$, indicating the 
RPT.  It commences from  $(T_z, P_z)$ and terminates at $(P_t, T_t$).
A similar figure is valid for any $d\geq 6$.  
}  
\label{Fig:PTZeroth1}
\end{center}
\end{figure}

In $d\geq 6$ the situation is more subtle and markedly different as shown in Figs. \ref{Fig:6DKerr} and \ref{Fig:6DKerrZorder}.  
For $P>P_c$, $G$  resembles the curve characteristic for the Schwarzschild-AdS black hole, known from the Hawking--Page transition \cite{HawkingPage:1983}: the upper branch corresponds to small unstable black holes with $C_P<0$ whereas the lower branch describes  stable large black holes with $C_P>0$.  There is a critical point at $P=P_c$, and for the range of pressures $P\in(P_t, P_c)$  and temperatures $T\in(T_t, T_c)$ there is a standard first order SBH/LBH phase transition, reminiscent of the Van der Waals phase transition  \cite{KubiznakMann:2012}. However for $P_c>P_z=P>P_t$ three separate phases of black holes emerge (see fig.~\ref{Fig:PTZeroth1}): intermediate black holes (IBH) (on the left), small  (middle), and large  (on the right). This holds for $T\in(T_t, T_z)$, $P\in (P_t, P_z)$ and  terminates at $T=T_t$. Small and large black holes are separated by a standard first order phase transition, but  the intermediate and small are separated by a finite jump in $G$, which in this range has a discontinuous global minimum (fig.~\ref{Fig:6DKerrZorder}).  This is the RPT, first observed in a nicotine/water mixture \cite{Hudson:1904}, and since seen in multicomponent fluid systems, gels, ferroelectrics, liquid crystals, and binary gases \cite{NarayananKumar:1994}.   Finally, for $T<T_t$ only one LBH phase  exists. 

Consider fig. \ref{Fig:6DKerrZorder}. If we start decreasing the temperature from, say $T=0.24$, the system follows the lower solid (red) curve until it joins the upper solid (red) curve---this corresponds to a first order SBH/LBH
phase transition. As $T$ continues to decrease the system follows this upper curve until  $T= T_0\in (T_t, T_z)$, where
$G$ has a discontinuity at its global minimum. Further decreasing $T$, the system jumps to the uppermost red line---this corresponds to the zeroth order phase transition between small and intermediate black holes.  

This novel situation is clearly illustrated in the $P-T$ diagrams in fig.~\ref{Fig:PTZeroth1}. 
There is the expected SBH/LBH line of coexistence corresponding to the liquid/gas Van der Walls case, ending in a 
critical point $(T_c,P_c)$.  This line terminates at $(T_t,P_t)$, where there is a ``triple point" between the small, intermediate, and large black holes. For smaller values of $T$ there is an unstable line of coexistence (not shown) between the IBHs and LBHs.  For 
$T\in(T_t, T_z)$ there is a new IBH/SBH line of coexistence (see inset of fig.~\ref{Fig:PTZeroth1})  that terminates in another ``critical point" $(T_z,P_z)$.  
The range for the RPT is quite narrow and must be determined numerically. For example  for $J=1$ and $d=6$ we obtain 
 $(T_t, T_z, T_c)\approx (0.2332, 0.2349, 0.3004)$ and
$(P_t, P_z, P_c) \approx (0.0553, 0.0579, 0.0958)$.
 
%\vspace{-0.7cm}

%\subsection{Equation of state}

\begin{figure}
\begin{center}
\rotatebox{0}{
\includegraphics[width=0.47\textwidth,height=0.32\textheight]{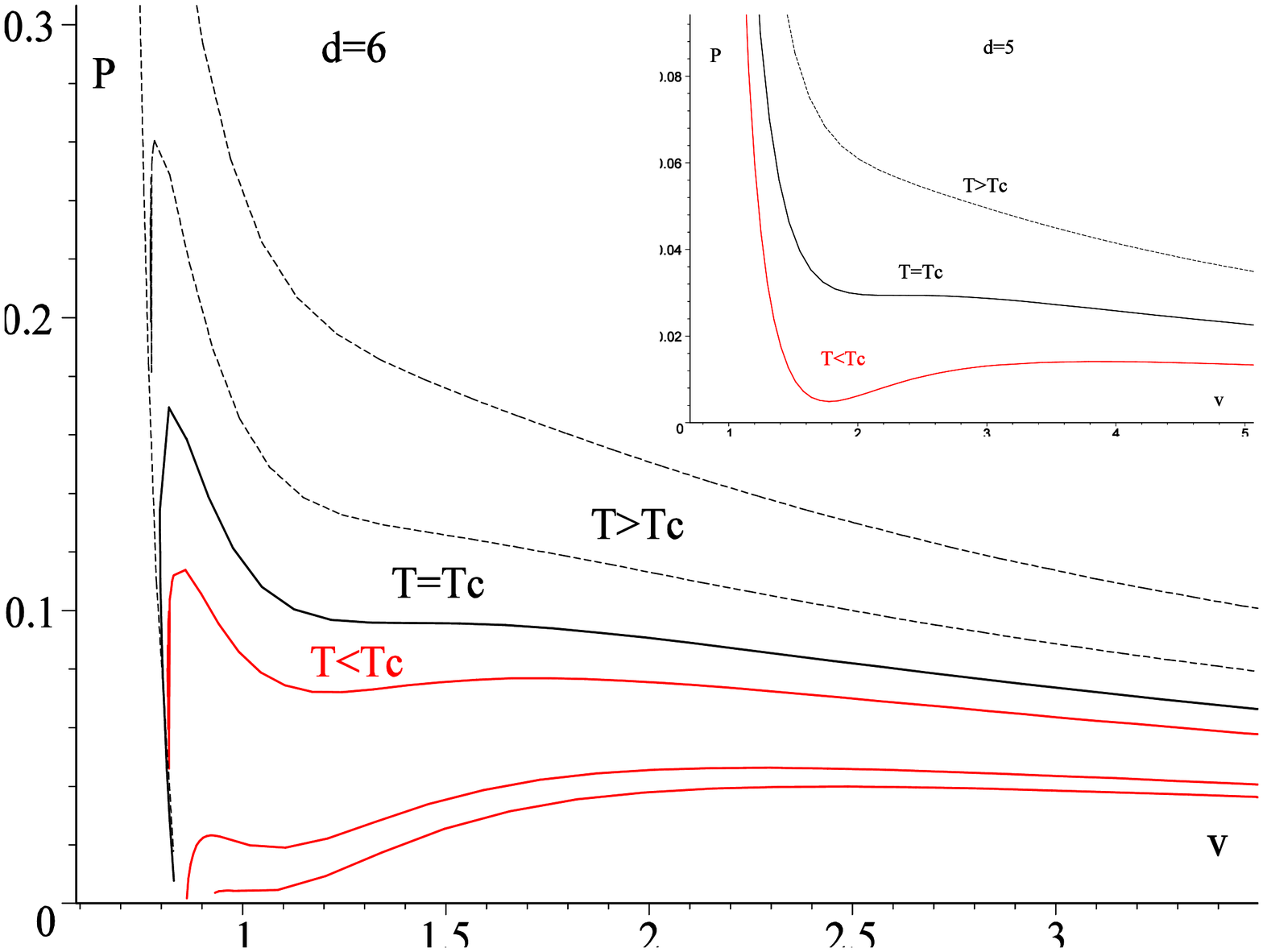}
}
\caption{{\bf $P-v$ diagram in $d=5$ (inset) and $d=6$.}
Obviously, in $d\geq 6$ the $P-v$ diagram is more complex than that of the standard Van der Waals, 
and reflects the interesting behavior of the Gibbs free energy and a possible RPT. 
}  
\label{Fig:6DKerrExact}
\end{center}
\end{figure}

{\em Equation of state.}
The equation of state $P=P(V,T,J)$ in the canonical (fixed $J$) ensemble can be computed by solving Eqs. \eqref{BHOM}---\eqref{BHS} to eliminate $(m,a,r_+)$ in terms of the basic thermodynamic variables. The result can be obtained
numerically (see fig.~\ref{Fig:6DKerrExact}), but there are  two cases of physical interest that can be approximated analytically: the slowly rotating ($a\to 0$)
case and the ultraspinning ($a\to l$) regime.

In the slowing rotating case, we expand in the parameter $\epsilon=a/l$, to obtain 
\be\label{Pbh}
P= \frac{T}{v}-\frac{d-3}{\pi(d-2)v^2}+\frac{\pi (d-1)16^d J^2}{4\omega_{d-2}^2 [(d\!-\!2) v]^{2(d\!-\!1)}} 
+O(\epsilon^4)
\ee
and it is straightforward to show that Van der Waals behaviour occurs as in $d=4$ \cite{GunasekaranEtal:2012}. The 
specific volume of the fluid $v$ is defined by $V=\frac{(\kappa v)^{d-1}\omega_{d-2}}{d-1}$, $\kappa=\frac{1}{4}(d-2)$.
Critical points ($P_c, v_c, T_c$) can be computed from $\frac{\partial P}{\partial v}=\frac{\partial ^2P}{\partial v^2}=0$. We find
that the critical exponents $\alpha=0$, $\beta=1/2$, $\gamma=1$, $\delta=3$ match those of a Van der Waals fluid,
though the critical ratio $\rho_c=\frac{P_c v_c}{T_c}=\frac{2d-3}{4(d-1)}$ 
differs; note that it reduces to $\rho_c=5/12$ for $d=4$  \cite{GunasekaranEtal:2012}.

In the ultraspinning limit $a\to l$, $r_+\to 0$.  For all $d\geq 6$
the geometry of a black hole approaches that of a black membrane \cite{EmparanMyers:2003, CaldarelliEtal:2008}; setting $f=1- \mu R^{5-d}$ the metric is
\be
ds^2_{M}=-fd\tau^2+\frac{d R^2}{f}+d\sigma^2+\sigma^2d\varphi^2+ R^2 d\Omega_{d-4}^2\,,\nonumber 
\ee 
known to be unstable due to the Gregory--Laflamme instability \cite{GregoryLaflamme:1993}. The entropy vanishes  and the temperature diverges but
both the angular momentum and (specific) volume $V=\frac{8\pi Ml^2}{(d-1)(d-2)}$ remain finite, while the equation of state is
$ P=\frac{4\pi}{(d-1)(d-2)}\frac{J^2}{V^2}$.
The same equation of state is valid for  ultraspinning black rings \cite{AltamiranoEtal:2013}. 
Expanding about the ultraspinning limit, $\Xi\to 0$, we find
\ba\label{Pultra}
P_6&=&\frac{\pi J^2}{5 V^2}-\frac{\sqrt{5\omega_4}}{16\sqrt{\pi^3 VT}}
+\frac{9\sqrt{5\omega_4}J^2}{3200\sqrt{\pi^3V^5T^5}}+ O(\Xi^9)\,, \nonumber\\
G_6&=&\frac{2\sqrt{5\pi P} J}{5}+\frac{\sqrt{25\omega_4 J}}{8\sqrt{T}(\pi P)^{\frac{1}{4}}}-\frac{5\omega_4}{1024 \pi^3 TP}
+O(\Xi^3)  \nonumber
\ea
for $d=6$; higher-dimensional expansions can likewise be computed. 

In between these two limiting cases a phase transition to a novel family of black objects branching off the spherical black holes (e.g., black rings) is expected (see, e.g., \cite{DiasEtal:2010, HartnettSantos:2013} and references therein).  The  exact critical point is shifted; we find numerically that the critical ratio $\rho_c$
is slightly smaller  and that all critical exponents remain the same.
\begin{figure}
\vspace{-0.7cm}
\begin{center}
\rotatebox{-90}{
\includegraphics[width=0.34\textwidth,height=0.31\textheight]{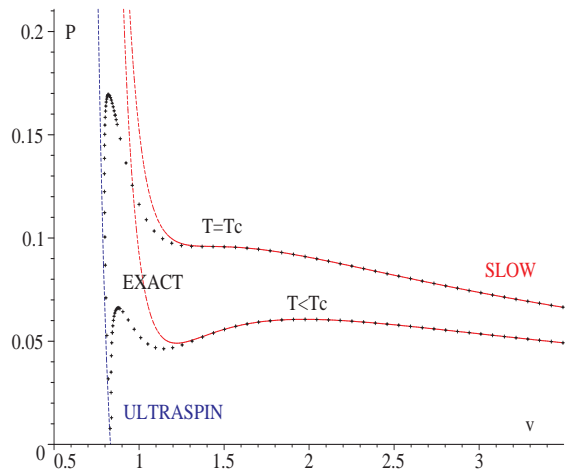}
}
\caption{{\bf The two approximations.}
Exact critical and subcritical isotherms, depicted by black crosses, are compared to the slow spinning expansion \eqref{Pbh} denoted by red curves and the ultraspinning expansion \eqref{Pultra} denoted by a blue curve. 
Note that the ultraspinning black holes correspond to the upper branch in the Gibbs free energy which is unstable. We have considered $d=6$ and set $J=1$.
}  
\label{Fig:6DKerr1st}
\end{center}
\end{figure} 
The two expansions for the equation of state are, together with the exact numerical solution, displayed for the critical temperature in fig.~\ref{Fig:6DKerr1st}.  Obviously, the slow rotation approximation is effectively accurate for large values of $r_+$ (rightmost curve) whereas the ultraspinning one (leftmost curve) is accurate as $r_+$ approaches zero.

%\section{Discussion}

{\em Discussion.}
Although the cosmological constant is generally regarded as fixed in the action, it is possible to dynamically generate it using a $(d-1)$-form gauge potential and incorporate it into a generalized first-law \cite{CreightonMann:1995}. We see that identifying it as pressure and incorporating its conjugate volume yields not only a consistent Smarr formula, but also a qualitatively new phase structure in the thermodynamics of rotating black holes similar to binary fluids. In binary fluids at low temperatures, directional bonding between unlike species can lead to a miscible state, which is restored at high temperatures since entropy of mixing dominates.  At intermediate temperatures the two fluids become immiscible.  The corresponding physics for $d\geq 6$ rotating black holes remains an interesting subject for further study.

%\bigskip
{\em Acknowledgments.} This work was supported in part by the Natural Sciences and Engineering Research Council of Canada.
We are grateful to the Perimeter Institute where part of this work was carried out.

\vspace{-0.75cm}

%%%%%%%%%%%%%%%%%%%%%%%%%%%%%%%%%%%%%%%%%%%%%%%%%%%%%%%%%%%%%%%%%%%%%%%%%
%%%%%%%%%%%%%%%%%%%%%%%%%%%%%%%%%%%%%%%%%%%%%%%%%%%%%%%%%%%%%%%%%%%%%%%%%

%\bibliography{Databaze}
%\bibliographystyle{JHEP}
%\end{document}

\providecommand{\href}[2]{#2}\begingroup\raggedright\endgroup

\end{document}